\documentclass[preprint2]{aastex62}

\usepackage{graphicx}
\usepackage{gensymb}
\usepackage[caption=false]{subfig}
\graphicspath{ {./figures/} }

\received{XXX}
\revised{XXX}
\accepted{XXX}
\submitjournal{ApJL}

\shorttitle{ZTF Companion Shocking}
\shortauthors{Burke et al.}

\begin{document}

\title{Companion Shocking Fits to the 2018 ZTF Sample of SNe Ia Are Consistent with Single-Degenerate Progenitor Systems}

\correspondingauthor{J. Burke (he, him)}
\email{jburke@lco.global}

\author[0000-0003-0035-6659]{J. Burke}
\affil{Las Cumbres Observatory, 6740 Cortona Dr, Suite 102, Goleta,
CA 93117-5575, USA}
\affil{Department of Physics, University of California, Santa Barbara, CA
93106-9530, USA}

\author[0000-0003-4253-656X]{D. A. Howell}
\affiliation{Las Cumbres Observatory, 6740 Cortona Dr, Suite 102, Goleta, CA 93117-5575, USA}
\affiliation{Department of Physics, University of California, Santa Barbara, CA 93106-9530, USA}

\author[0000-0003-4102-380X]{D. J. Sand}
\affil{Steward Observatory, University of Arizona, 933 North Cherry Avenue, Tucson, AZ 85721-0065, USA}

\author[0000-0002-0832-2974]{G. Hosseinzadeh}
\affil{Steward Observatory, University of Arizona, 933 North Cherry Avenue, Tucson, AZ 85721-0065, USA}

\begin{abstract}
\noindent The early lightcurves of Type Ia supernovae (SNe Ia) can be used to test predictions about their progenitor systems.
If the progenitor system consists of a single white dwarf in a binary with a Roche-lobe-overflowing non-degenerate stellar companion,
then the SN ejecta should collide with that companion soon after the explosion and get shock-heated,
leaving an early UV excess in the lightcurve.
This excess would only be observable for events with favorable viewing angles,
$\sim$10\% of the time.
We model the 2018 ZTF sample of 127 SNe Ia using companion shocking models,
and recover an observed early excess rate of $12.0\pm3.6\%$,
consistent both with several other rates calculated throughout the literature,
and with the expectation that SNe Ia predominantly occur in single-degenerate systems.
We observe early excesses only in spectroscopically normal SNe Ia,
in contradiction to the claim that such excesses occur more frequently in overluminous SNe Ia.
We also show that the detection of early excesses can be methodology-dependent.
We encourage the observation of large samples of SNe Ia with high-cadence multiwavelength early data in order to test the statistical predictions of SN Ia progenitor models,
and we also encourage the refinement of existing models.

\end{abstract}


\section{Introduction} \label{sec:intro}

Type Ia supernovae (SNe Ia) come from exploding white dwarfs (WDs).
This statement is uncontroversial and has been understood for decades \citep{hoyle},
but almost every detail of the progenitor system and explosion mechanism is the subject of active research.
What mass are the WDs when they explode?
Do they need to accrete mass up to the Chandrasekhar limit,
or, as some models predict \citep[e.g.][]{polin_double_det, shen_subchandra},
can they explode at sub-Chandrasekhar masses?
Where does the explosion start?
Does it begin roughly at the center of the WD \citep{khokhlov91},
or does it begin in a surface layer of accreted He which causes the underlying WD to detonate \citep{polin_double_det}?
How do the WDs gain enough mass to explode?
Are they in a binary system with a Roche-lobe-overflowing nondegenerate stellar companion \citep[referred to as the ``single-degenerate" case;][]{hoyle},
or does the primary WD tidally disrupt a less massive secondary WD \citep[the ``double-degenerate" case;][]{iben}?
These and many other questions have yet to be definitively answered.

SNe Ia have standardizable lightcurves, especially around peak
\citep[see e.g.][]{riessnobel,perlmutter_99_stretch,Phillips99}.
But their early lightcurves,
within a few days of explosion,
are much less homogeneous and can contain observational signatures which reveal information about their progenitor systems.
One such signature was predicted in \citet{kasen}:
in the single-degenerate case,
as the SN ejecta collide with a nondegenerate companion they will get shock-heated,
resulting in early UV excesses which should be observable for binaries with favorable viewing angles ($\sim$10\% of events).
After these early UV excesses were predicted they (or similar effects) have subsequently been observed in a small number of objects:
SN 2012cg \citep{marion},
iPTF14atg \citep{cao},
SN 2016jhr \citep[aka MUSSES1604D,][]{jiang_1604D},
iPTF16abc \citep{miller_16abc},
SN 2017cbv \citep{griffin},
SNe 2017erp and 2018yu \citep{burke_LCO_sample},
SN 2019yvq \citep{miller_19yvq,siebert_19yvq,tucker_19yvq,burke_19yvq},
and
SN 2021aefx \citep{ashall_2021aefx,griffin_2021aefx}.
Some other objects have early excesses,
but without the color information needed to determine their temperature,
such as
SN 2018oh \citep{wenxiong_18oh,dimitriadis_18oh,shappee_18oh}
and
SN 2020hvf \citep{jiang_20hvf},
and still others 
\citep[SN 2018aoz;][]{ni_18aoz, ni_18aoz_2}
have early color evolution which can differ by more than a magnitude from a typical SN Ia.

Most of the above objects were modeled with companion interaction models,
but there are other physical models which can produce early excesses.
``Double-detonation" models,
where the WD builds up a layer of He on its surface until the He detonates,
driving a shockwave into the WD causing it to detonate in turn,
can also produce a range of early lightcurve behavior due to the presence of extra radioactive products in the outer ejecta \citep[see][]{sim_double_det,polin_double_det}.
Models which vary the distribution of $^{56}$Ni,
which powers SN Ia lightcurves \citep{pankey_dissertation},
can also produce a range of early behavior,
including early bumps \citep{turtls,magee_maguire,magee_models}.
Both classes of models produce extra radioactive material in the outer ejecta,
resulting in some similar effects (e.g. ``red bumps" at early times),
which makes them potentially difficult to distinguish for near-Chandrasekhar-mass WDs.

Recently,
\citet{burke_LCO_sample} examined a sample of 9 SNe Ia with exemplary high-cadence multiwavelength early data.
Overall the paper favored companion interaction models to explain both objects which exhibit an early excess,
and those that do not.
The sample was constructed using a set of criteria to make it as unbiased as possible,
and contained one object with a strong early excess and two others with weaker excesses.
Based on the distributions of early excess strengths and best-fit viewing angles,
that paper concluded that there was not enough evidence to disprove the null hypothesis,
i.e. that all SNe Ia come from single-degenerate progenitor systems.

In this paper we will be focusing solely on companion shocking models to explain early excesses.
We do not claim that the models can perfectly explain every aspect of the dataset:
we refer to the discussion section of \citet{burke_LCO_sample} for an explicit list of their pros and cons.
Perhaps the strongest point against them is
the lack of H observed in the nebular spectra of SNe Ia \citep[see e.g.][]{dave_nebular_halpha, tucker_nebular_halpha},
even though the ejecta-companion interaction should strip H from a non-degenerate companion \citep{botyanski_halpha,dessart_nebular_H}.
They also struggle to fit UV data:
even though the process of shock heating should make the early excess strongest in the UV ,
the models consistently overpredict the UV flux \citep[see e.g.][]{griffin,griffin_2021aefx}.
Keeping these caveats in mind \citep[and again, for a more in-depth discussion we refer to][]{burke_LCO_sample},
we nevertheless favor the models and use them to investigate the sample of SNe Ia presented here.

This paper follows very similar methodology to \citet{burke_LCO_sample},
but applied to a different sample of SNe Ia.
The sample here has an order of magnitude more objects but limited multi-band information when compared to \citet{burke_LCO_sample},
making this a complementary analysis.
In Section \ref{sec:data_and_sample} we briefly describe our data and sample,
and in Section \ref{sec:model} we describe the companion interaction models we use to fit the data.
We show and discuss the results of the fits in Section \ref{sec:results},
before concluding in Section \ref{sec:conclusion}.

\section{Data and Sample} \label{sec:data_and_sample}

The full dataset in this paper comes from \citet{yao_2019},
so we refer to that paper for details of its acquisition and reduction.
The data come from the ZTF survey \citep{bellm_ztf},
and the dataset has many advantages:
the large number of objects (127);
the uniform inclusion of non-detections;
and the high cadence of coverage,
with some objects being observed not only nightly, but revisited six times per night
(thrice in $g$ and thrice in $r$).
The primary downside of the dataset is that it is in only two filters,
$g$ and $r$,
which strongly limits the number of filter combinations which can be used to determine color/temperature 
\citep[compare e.g. the $UBVgri$ data used in][]{burke_LCO_sample}.

To obtain the data,
we simply downloaded the digital version of Table 5 of \citet{yao_2019},
which is a FITS file containing the full dataset in flux space.
We convert to magnitude space following Equations 7--10 of that paper.

The sample was introduced in a series of three papers:
Paper I \citep{yao_2019} described the overall data reduction method and the properties of the sample itself;
Paper II \citep{miller_rise_times} modeled the rising lightcurves of the sample to infer rise times;
and Paper III \citep{ztf_colors} analyzed the sample's color evolution.
Again,
we refer to \citet{yao_2019} for a full description of the sample.
In short,
the sample contains 127 SNe Ia.
All objects have observations earlier than at least 10 rest-frame days before maximum light,
and 50 of the objects have detections earlier than 14 rest-frame days before peak.
The median redshift of the sample is 0.076.
Though we focus on photometry here,
\citet{yao_2019} also discusses the spectroscopic classifications of the objects:
107 objects are ``normal" SNe Ia
(including the 25 SNe listed in the paper as ``normal*," i.e. tentatively classified as normal),
10 objects are 99aa-likes
(including three tentative classifications),
four objects are super-Chandrasekhar SNe Ia
(including two tentative classifications),
three are 91T-likes
(including two tentative classifications),
one is an 86G-like,
one is an 02cx-like,
and one is a Ia-CSM.
We again refer to \citet{yao_2019} for details of the classifications,
and also to \citet{taubenberger_SN_handbook} for a discussion of the various different subtypes of SNe Ia.

\section{Models} \label{sec:model}

\subsection{Initial Parameter Measurements} \label{ssec:snoopy}

Following \citet{burke_LCO_sample},
we utilize the Python package \texttt{SNooPy} \citep{snoopy} to measure several necessary quantities for the objects.
We do \texttt{SNooPy} fits using the default \texttt{EBV\_model} and the \texttt{fitMCMC()} procedure,
enforcing $R_{V,\rm{host}} = 3.1$.
We impose a uniform prior on \texttt{EBVhost} ranging from 0 to 1.
We limit the data to the epochs relevant for \texttt{SNooPy} models,
i.e. $-10$ to $+50$ days from peak \citep[using $\rm{MJD}_{\rm{peak}}$ as measured by][]{yao_2019}.
We visually inspect fits to ensure that they are reasonable.
We adopt the \texttt{SNooPy} values of distance moduli and host extinction for all objects throughout our analysis.
We convert each object's $E(B-V)$ (Milky Way $+$ host) to per-filter extinction values
using the \citet{schlafly_finkbeiner} recalibration of the \citet{schlegel_dust_map} dust maps,
accessed by the Python package \texttt{extinction} \citep{extinction_zenodo}.

We also use \texttt{SNooPy} to measure K-corrections for the data,
similar to \citet{ztf_colors}.
The way that \texttt{SNooPy} fits SN Ia lightcurves already involves K-corrections,
making them straightforward to extract from the fits.
This method of K-correction is the largest deviation from the methodology of \citet{burke_LCO_sample},
which otherwise this paper follows quite closely.
The sample of \citet{burke_LCO_sample} was low redshift ($z<0.01$),
so any K-correction would both be small and
(as is inherent in doing K-corrections)
would rely on assumptions about the underlying SED,
even though those assumptions could be incorrect for objects with early excesses
since their early spectra differ from SNe Ia without excesses
\citep[see e.g.][]{marion, griffin, burke_19yvq}.
Due to the much higher redshift of the sample here
(median redshift 0.076)
we deemed K-corrections necessary,
since the models described in Section \ref{ssec:cs_models} rely on accurate absolute magnitudes both to compare to the rest-frame $g$ and $r$ template lightcurves
and to compare to the strengths of any early excesses.

\subsection{Companion Shocking Models} \label{ssec:cs_models}

\subsubsection{Description of models} \label{sssec:cs_models_description}

\begin{figure*}[t!]
\begin{center}
\includegraphics[width=0.98\textwidth]{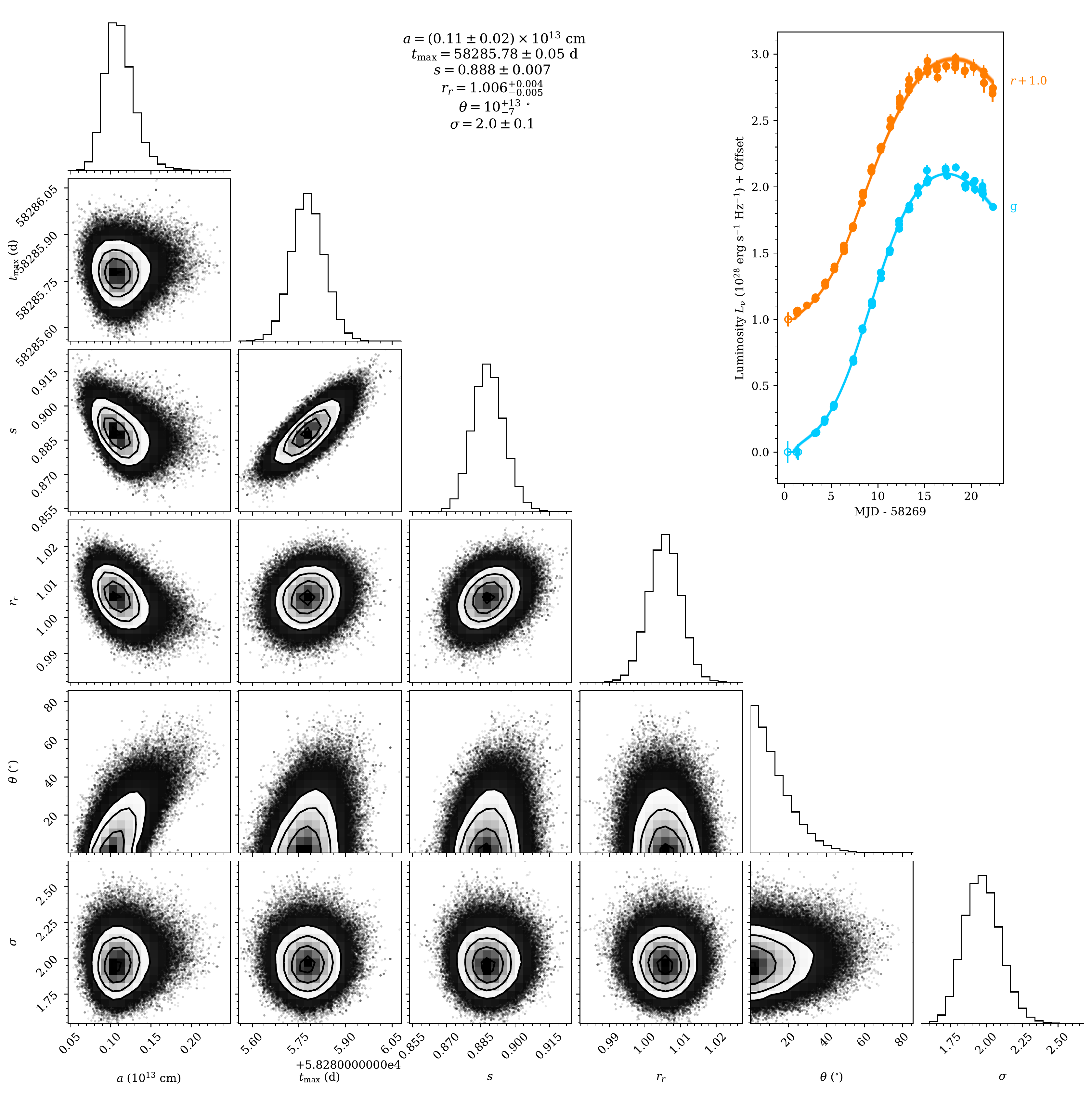}
\caption{
Corner plot of the model described in Section \ref{ssec:cs_models} for ZTF18aaxsioa,
one of the objects we identify as having an early excess due to the fact that its posterior distribution of the $\theta$ parameter is peaked to $\theta=0\degree$.
The inset data show the extremely high cadence and useful non-detections 
(first epoch shown, unfilled circles)
typical of the sample.
}
\label{fig:corner}
\end{center}
\end{figure*}

Our models are similar to the ones used in \citet{burke_LCO_sample} 
\citetext{\citealp[along with][]{griffin}
\citealp[and][and to some extent those in]{griffin_2021aefx}
\citealp[][]{dimitriadis_18oh}
\citealp[and][]{miller_19yvq}}.
They make use of the \texttt{lightcurve\_fitting} Python package \citep{griffin_lightcurvefitting},
which performs MCMC fits using the \texttt{emcee} package \citep{emcee}.
The model consists of two components:
a template lightcurve from \texttt{SiFTO} \citep{sifto}
which is scaled and stretched to maximize overlap with the data,
to which is added a blackbody component (which can dominate at early times) representing a companion shock interaction based on the analytical formulae from \citet{kasen}.
As in \citet{burke_LCO_sample}
we add a parameter to represent the viewing angle,
implemented as a multiplicative factor on the shock component following the semi-analytic formulation of \citet{brown_2012}.

This results in a total of six parameters:
\begin{enumerate}
    \item $a$, the companion separation of the shock component
    \item $t_{\rm{max}}$, the time of $B$-band maximum light for the stretch component
    \item $s$, the stretch applied to the stretch component
    \item $r_{r}$, a factor on the $r$-band flux of the stretch component
    \item $\theta$, the viewing angle (which determines a multiplicative factor on the shock component)
    \item $\sigma$, a multiplicative factor on the errors of the data to account for error underestimation.
\end{enumerate}
The models are identical to those used in \citet{burke_LCO_sample},
except for the fact that two parameters used in that paper are excluded because they affect only non-$gr$ data and are thus irrelevant for the dataset here.
We refer to that paper for a more detailed description of the models -- the two differences here are the slightly fewer parameters and the fact that we do the fits on the K-corrected data.
We also refer to the discussion section of that paper for an explicit list of the pros and cons of the model,
including various simplifying assumptions it makes.

A corner plot for an object we identify as having an early excess is shown in Figure \ref{fig:corner}.

\subsubsection{Is It Valid to Use Just $gr$ Data in Companion Shocking Models?} \label{sssec:gr}

\begin{figure}[t!]
\begin{center}
\includegraphics[width=0.47\textwidth]{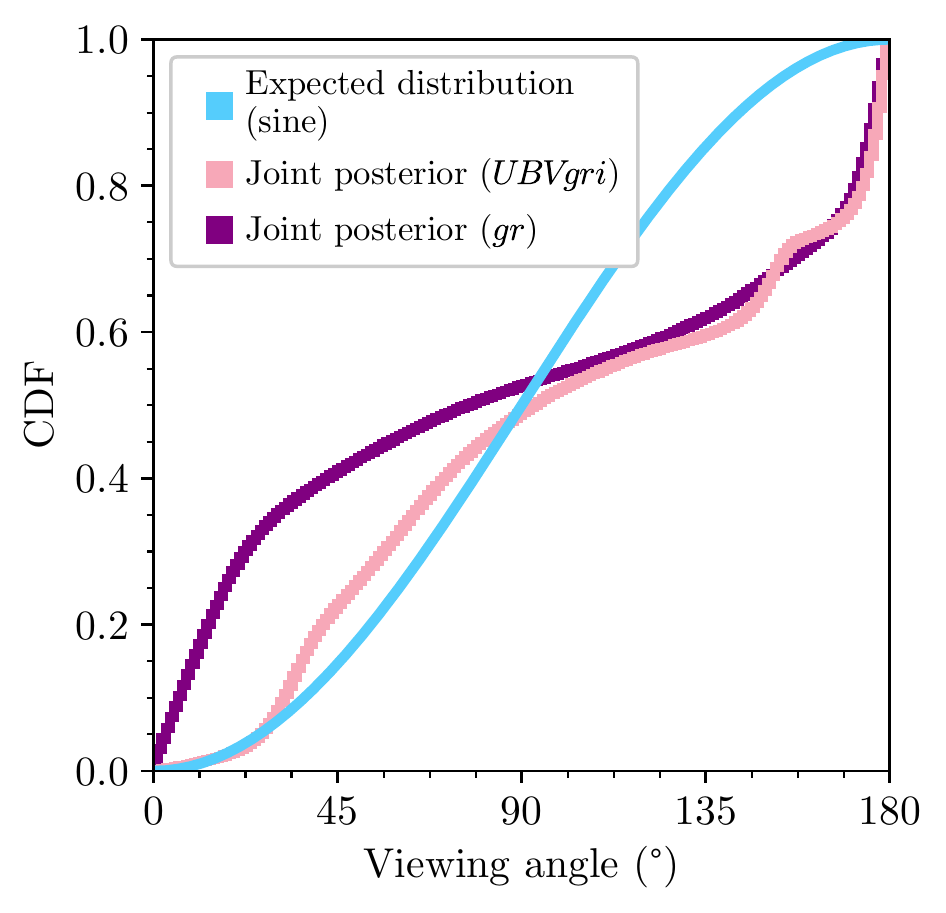}
\caption{
Joint viewing angle posteriors for the sample modeled in \citet{burke_LCO_sample},
which markedly change from the result presented in that paper 
(the pink line)
when the lightcurves are limited to just $gr$ data (the purple line).
We conclude that these models can still be used to detect early excesses in $gr$ data,
as objects exhibiting early excesses converge to low viewing angles,
but the models shouldn't be used to measure physical parameters.
It also means that the K-S test method used in \citet{burke_LCO_sample} is no longer applicable for $gr$ data alone.
}
\label{fig:LCO_theta}
\end{center}
\end{figure}

\begin{figure}[t!]
\begin{center}
\includegraphics[width=0.47\textwidth]{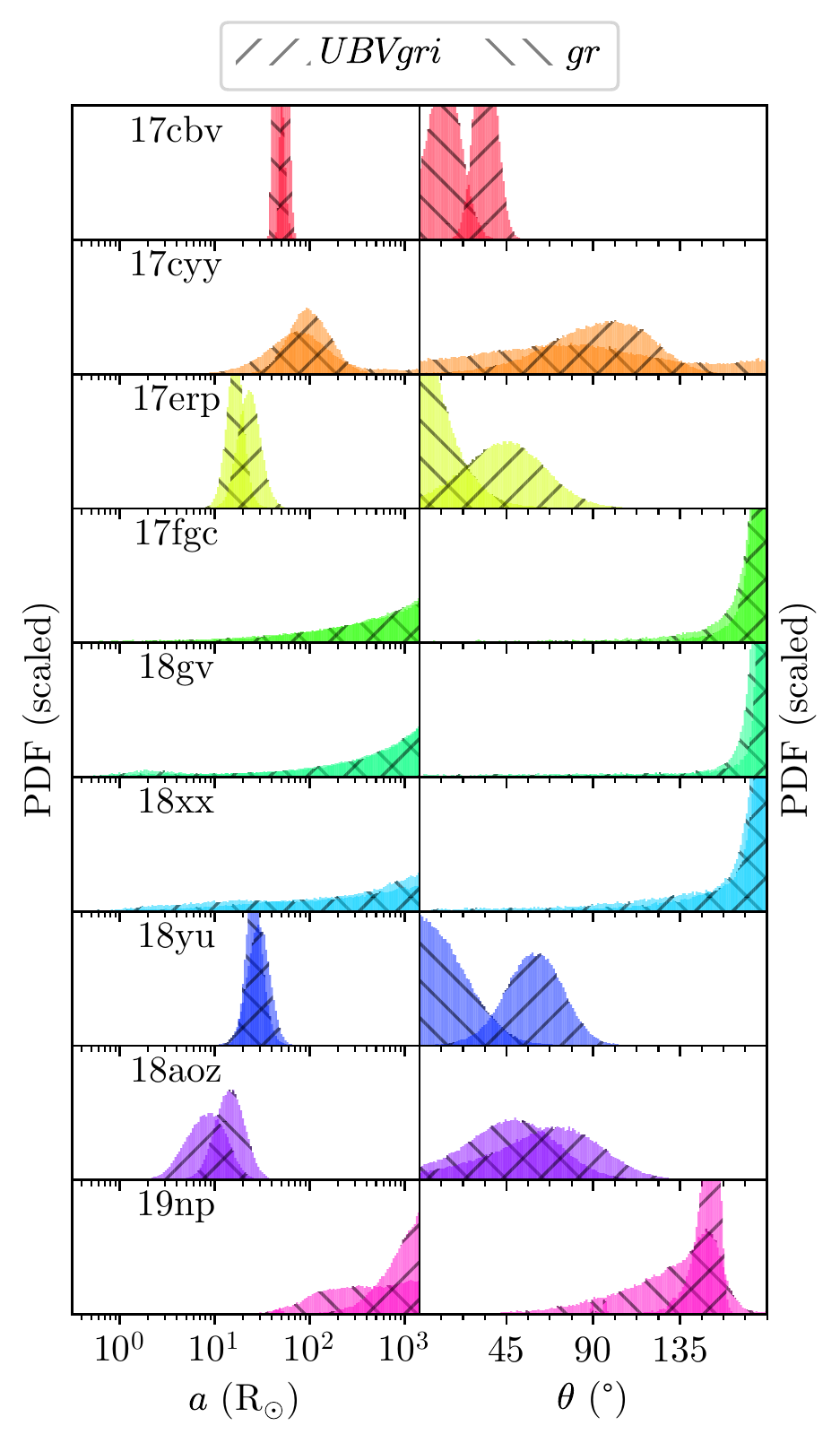}
\caption{
The per-object posteriors of the sample presented in \citet{burke_LCO_sample},
which change depending on whether we use the object's $UBVgri$ dataset 
(as done in that paper)
or just its $gr$ data
(as we do for the sample presented here).
The three objects identified in that paper as having early excesses
(SNe 2017cbv, 2017erp, and 2018yu)
have best-fit viewing angles which shift towards $\theta=0$ when the data are limited to $gr$.
}
\label{fig:posterior_shifts}
\end{center}
\end{figure}

Most of the literature using this style of modeling has access to highly multiwavelength datasets
\citetext{\citealp[see][]{burke_LCO_sample}
\citealp[for the most direct comparison, but also][]{cao, griffin,dimitriadis_18oh,miller_19yvq,burke_19yvq,griffin_2021aefx}}.
The wavelength coverage is often $UBVgri$,
but sometimes extends further into the UV with data from the
Neil Gehrels Swift Observatory \citep[{Swift};][]{Gehrels04}.
The temperature of the early data,
accessed through different filter/color combinations,
is a critical parameter implicitly measured in the models,
making it natural to ask:
is it valid to use only $gr$ data with these models?

To summarize the answer to that question:
these models can still detect early excesses using only $gr$ data,
but they should not be used to measure physical parameters.

To see why this is the case,
we reexamine the objects modeled in \citet{burke_LCO_sample},
limiting the data from $UBVgri$ down to $gr$.
One result from that paper had to do with the joint MCMC posteriors of the $\theta$ parameter from the nine objects in that sample,
where,
when comparing the joint posterior to the expected distribution of viewing angles if all SNe Ia arose from single-degenerate systems,
there was not enough evidence to disprove that null hypothesis.

Figure \ref{fig:LCO_theta} shows the joint viewing angle posterior both as it was presented in that paper (the pink line),
and when the data are limited to just $gr$ (the purple line).
This behavior is shown per-object in Figure \ref{fig:posterior_shifts},
for both the $\theta$ and $a$ posteriors.
Looking at the figures,
it's clear that objects which have no detectable excess have still converged to having their maximum likelihood at $\theta=180\degree$,
as they did in the $UBVgri$ models
(see SNe 2017fgc, 2018gv, and 2018xx in Figure \ref{fig:posterior_shifts}).
However,
the three objects with detectable excesses
(SNe 2017cbv, 2017erp, and 2018yu)
now have a maximum likelihood closer to $\theta=0\degree$ in the $gr$ fits.

\begin{deluxetable*}{cccccccccc}
\tablehead{\colhead{ZTF18} & \colhead{$a\textrm{ (R}_{\odot}\textrm{)}$} & \colhead{$t_{\rm{max}}$ (MJD)} & \colhead{$s$} & \colhead{$r_{r}$} & \colhead{$\theta$ $(\degree)$} & \colhead{$\sigma$} & \colhead{IAU} & \colhead{z} & \colhead{Subtype}}
\startdata
\multicolumn{10}{c}{Gold tier} \\
\hline
aaxsioa & $15.8^{+2.7}_{-2.3}$ & $58285.78^{+0.05}_{-0.05}$ & $0.888^{+0.007}_{-0.007}$ & $1.006^{+0.004}_{-0.005}$ & $10^{+13}_{-7}$ & $1.97^{+0.13}_{-0.12}$ & cfa & 0.0315 & normal* \\
abcflnz & $9.4^{+1.5}_{-1.4}$ & $58304.78^{+0.08}_{-0.08}$ & $0.962^{+0.008}_{-0.007}$ & $0.969^{+0.004}_{-0.004}$ & $6^{+9}_{-5}$ & $2.03^{+0.08}_{-0.07}$ & cuw & 0.0273 & normal \\
\multicolumn{10}{c}{Silver tier} \\
\hline
abssuxz & $34^{+17}_{-11}$ & $58376.82^{+0.20}_{-0.19}$ & $0.860^{+0.024}_{-0.018}$ & $0.987^{+0.014}_{-0.014}$ & $31^{+47}_{-23}$ & $1.15^{+0.13}_{-0.11}$ & gfe & 0.0649 & normal \\
abxxssh & $47^{+28}_{-15}$ & $58396.54^{+0.22}_{-0.19}$ & $1.077^{+0.027}_{-0.022}$ & $0.902^{+0.013}_{-0.014}$ & $27^{+34}_{-19}$ & $1.49^{+0.12}_{-0.11}$ & gvj & 0.0782 & normal \\
aavrwhu & $31.5^{+8.7}_{-7.5}$ & $58275.29^{+0.13}_{-0.13}$ & $1.090^{+0.019}_{-0.017}$ & $0.967^{+0.006}_{-0.007}$ & $15^{+22}_{-11}$ & $1.42^{+0.13}_{-0.11}$ & bxo & 0.062 & normal \\
abfhryc & $15.9^{+6.9}_{-4.2}$ & $58322.39^{+0.09}_{-0.08}$ & $0.980^{+0.010}_{-0.009}$ & $1.009^{+0.005}_{-0.005}$ & $25^{+27}_{-18}$ & $1.93^{+0.13}_{-0.12}$ & dhw & 0.0323 & normal \\
\multicolumn{10}{c}{Bronze tier} \\
\hline
aawjywv & $38^{+19}_{-12}$ & $58270.85^{+0.18}_{-0.17}$ & $0.855^{+0.024}_{-0.020}$ & $0.962^{+0.015}_{-0.016}$ & $30^{+40}_{-22}$ & $2.02^{+0.19}_{-0.16}$ & ccj & 0.0509 & normal* \\
aaqcozd & $170^{+67}_{-25}$ & $58253.02^{+0.11}_{-0.11}$ & $0.825^{+0.020}_{-0.015}$ & $0.863^{+0.019}_{-0.015}$ & $37^{+40}_{-25}$ & $0.92^{+0.09}_{-0.08}$ & bjc & 0.0732 & normal \\
abdfazk & $94^{+53}_{-23}$ & $58306.60^{+0.22}_{-0.20}$ & $0.906^{+0.032}_{-0.025}$ & $0.924^{+0.019}_{-0.017}$ & $43^{+49}_{-29}$ & $1.17^{+0.11}_{-0.10}$ & dbe & 0.084 & normal \\
abimsyv & $55^{+28}_{-19}$ & $58333.03^{+0.19}_{-0.18}$ & $1.003^{+0.021}_{-0.020}$ & $0.915^{+0.012}_{-0.012}$ & $41^{+31}_{-26}$ & $1.16^{+0.07}_{-0.06}$ & eni & 0.088 & normal* \\
aazsabq & $12.1^{+4.8}_{-3.3}$ & $58293.40^{+0.10}_{-0.10}$ & $0.891^{+0.012}_{-0.011}$ & $1.006^{+0.006}_{-0.006}$ & $21^{+26}_{-15}$ & $1.38^{+0.09}_{-0.08}$ & crn & 0.06 & normal \\
\enddata
\caption{
Values of the best-fit parameters and some object properties for the 11 SNe we identify as having early excesses.
The ``ZTF18" column is the ZTF name (with the ``ZTF18-" excluded for space).
The next six columns are the best-fit parameters from the models described in Section \ref{ssec:cs_models}.
Lastly we give the IAU name
(again excluding the ``SN 2018-" for space),
the redshift,
and the classification as listed in \citet{yao_2019}.
We have split the objects into tiers based on visual inspection of the fits and their residuals
(see Figure \ref{fig:residuals_EEx} and surrounding discussion).
}
\label{tab:EEx_parameters}
\end{deluxetable*}

This has three implications when the data are limited to $gr$.
One is that the best-fit values of the two relevant physical parameters
(i.e. $\theta$ and $a$)
should not fully be believed:
for example,
even though the viewing angles seem to have converged nicely for the $gr$ models,
there is a systematic offset from the best-fit $UBVgri$ values for the three objects with early excesses
(SNe 2017cbv, 2017erp, and 2018yu).
Their best-fit companion separations are also systematically lower,
though at lower significance.
However,
even though the $gr$ models shouldn't be used to measure physical parameters,
they can still be used to detect early excesses.
Early excesses are still present in the data,
regardless of filter 
\citep[see Figure 3 of][]{burke_LCO_sample},
and will manifest as $\theta$ posteriors which peak towards low viewing angles.
Lastly,
it also means that the K-S test done in \citet{burke_LCO_sample} to compare the joint viewing angle posterior to the null hypothesis is no longer applicable,
since the per-object posteriors have changed significantly.

\section{Results} \label{sec:results}

\subsection{Early Excess Rate}\label{ssec:EEx_rate}

In \citet{burke_LCO_sample},
SNe were classified as having an early excess if they met three criteria:
(1) if the object had data within five rest-frame days of inferred first light
\citep[the typical epochs for an early excess in][]{kasen},
(2) if the best-fit viewing angle was less than 90$\degree$,
and (3) if the residuals with respect to the stretch template showed a systematic which was representative of an early-but-fading shock component
(i.e. if they showed an initial discrepancy of $>$5$\sigma$ which decreased over time).
We use similar criteria here:
we keep criterion (1) unchanged;
following the results of Section \ref{sssec:gr} we change criterion (2) to a best-fit $\theta\leq45\degree$;
and we loosen the requirement of criterion (3) due to the lower S/N of the dataset,
although such a pattern is still obvious for some objects.

Using the above criteria,
we identify 11 objects with early excesses in our sample.
Table \ref{tab:EEx_parameters} lists the best-fit parameters and some object properties for these 11 SNe.
We have split the objects into tiers based on visual inspection of their fits and residuals
(see Figure \ref{fig:residuals_EEx} and surrounding discussion).
Since the sample contains 127 SNe in total,
these 11 objects na\"{i}vely represents 8.7\% of the total sample,
although as we will discuss more shortly,
the intrinsic rate of early excesses is slightly higher due to the low S/N of early data potentially obscuring some excesses.

\begin{figure}[t!]
\begin{center}
\includegraphics[width=0.47\textwidth]{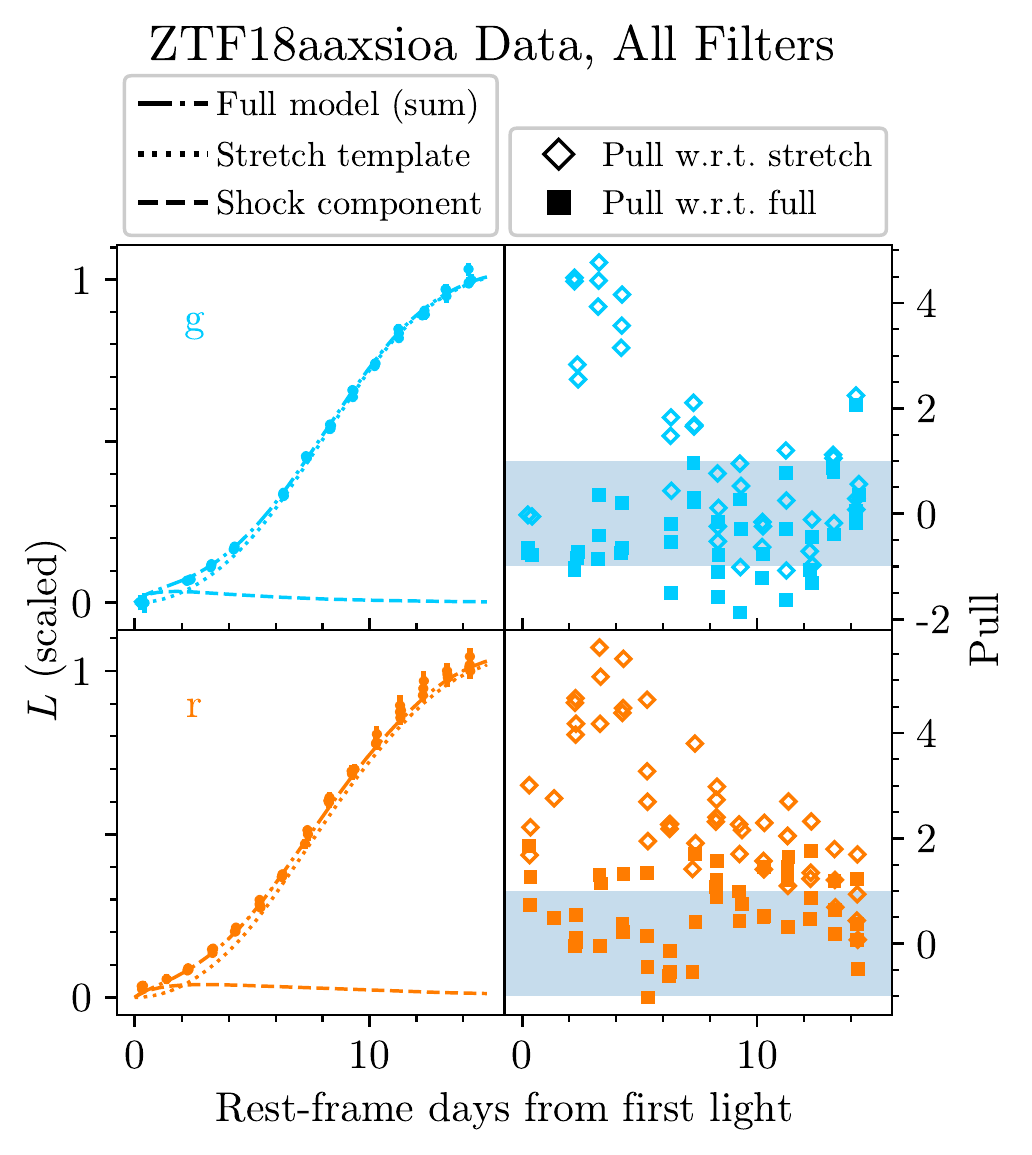}
\caption{
The best-fit model and residuals for ZTF18aaxsioa,
an object with an early excess.
The left panels show the per-filter data and the two components of the model:
the stretched template (dotted line) and the shock component (dashed),
which are added together to the full model.
The right panels show the residuals with respect to both the full model 
(filled squares)
and the stretched template 
(empty diamonds),
where the signature of the early excess is obvious (initial discrepancy which vanishes over time).
``Pull" is defined as the number of standard deviations a data point is away from the relevant model,
divided by the best-fit $\sigma$ value 
-- the majority of square points should cluster within one standard deviation of zero (the grey shaded box), which they do.}
\label{fig:residuals}
\end{center}
\end{figure}

Figure \ref{fig:residuals} shows the best-fit model and residuals for a single object which we identify as having an early excess
(ZTF18aaxsioa, the same object shown in Figure \ref{fig:corner}).
The distinctive pattern in the residuals with respect to the stretch template (diamonds),
where their discrepancy diminishes over time,
is characteristic of an early-but-fading shock component 
\citep[see Figure 3 of][]{burke_LCO_sample}.

The sample modeled in \citet{burke_LCO_sample} consisted of nine objects,
three of which had early excesses:
that paper therefore calculated an early excess rate from simple Poissonian statistics of $33\pm19\%$.
However,
the objects in that paper were extremely low-redshift ($z<0.01$),
and all objects were detected well above the telescopes' limiting magnitudes within five days of inferred first light.
\citep[``First light" in this context being the time when the SN flux is first detectable,
which is not necessarily the same as when the explosion occurred due to ``dark phases" in SNe Ia when the radiation has not yet diffused out of the ejecta -- see][where this is discussed in more detail.] {piro_dark_phase}
\citet{burke_LCO_sample} assumed that any potential early excesses would be revealed by the data.
The sample here is significantly different:
it is much higher redshift (median redshift of 0.076),
resulting in lower average signal-to-noise (S/N) in the crucial earliest epochs.


\begin{deluxetable}{cccc}
\tablehead{\colhead{S/N cut ($\sigma$)} & \colhead{$n_{\rm{objects}}$} & \colhead{$n_{\rm{EEx}}$} & \colhead{Rate}}
\startdata
N/A & 127 & 11 & $8.7 \pm 2.6 \%$ \\
3 & 103 & 11 & $10.7 \pm 3.2 \%$ \\
5 & 92 & 11 & $12.0 \pm 3.6 \%$ \\
8 & 68 & 11 & $16.2 \pm 4.9 \%$ \\
10 & 49 & 9 & $18.4 \pm 6.1 \%$ \\
15 & 26 & 5 & $19.2 \pm 8.6 \%$ \\
20 & 18 & 3 & $16.7 \pm 9.6 \%$ \\
30 & 9 & 2 & $22 \pm 16 \%$ \\
50 & 4 & 2 & $50 \pm 35 \%$
\enddata
\caption{
The effect on the early excess (EEx) rate of imposing different S/N cuts on the early data,
e.g. requiring that at least one epoch 
within five rest-frame days of inferred first light
has a $>$10$\sigma$ detection.
The two early-excess SNe that survive the $50\sigma$ cut are the two gold-tier objects 
(see Table \ref{tab:EEx_parameters}).
We quote an overall early excess rate of $12.0 \pm 3.6 \%$,
the value from the $5\sigma$ cut,
since the rates calculated for more stringent S/N cuts are all consistent 
(although systematically higher)
than that value.
}
\label{tab:StN_cuts}
\end{deluxetable}

This lower S/N could serve to obscure weak early excesses hiding in the data.
For instance,
in this sample only 81\% of the objects (103 out of 127) are detected at all within 5 rest-frame days of inferred first light.
Table \ref{tab:StN_cuts} shows the results of imposing different S/N cuts on the early data of the sample,
e.g. only considering objects which have a $>$10$\sigma$ detection with 5 days of first light.
Each S/N cut leaves some total number of objects,
and also some subset of the early-excess objects,
from which an early excess rate can be calculated.
The $30\sigma$ cut leaves a sample extremely similar to the one in \citet{burke_LCO_sample}:
that paper had an early excess rate of 
3 out of 9 objects ($33\pm19\%$),
and limiting the sample here to SNe with a $30\sigma$ detection in the epochs where an early excess could be detected yields a rate of
2 out of 9 objects ($22\pm16\%$).
We thus believe that the sample here is identical to the sample of \citet{burke_LCO_sample},
except at higher redshift and observed in fewer bands.
As in that paper,
the distribution of early excess strengths here is consistent with SNe Ia predominantly arising in single-degenerate systems.

Which rate should we quote as the single rate of early excesses in SNe Ia?
As is obvious in Table \ref{tab:StN_cuts},
using a less restrictive S/N cut results in a smaller formal uncertainty due to the larger sample size,
but it also means that lower intensity early excesses could be hidden in the data, undetected.
We use the value calculated from the S/N$>$5 cut,
since that S/N is sufficient to detect strong early excesses,
and since rates calculated with higher S/N cuts are all consistent
(although systematically higher)
than that value.
It is possible that the higher rates derived from stricter S/N cuts are consistent with the prediction in \citet{kasen},
i.e. that higher S/N data could detect weaker excesses,
leading to a higher overall rate of excesses,
but the uncertainties on the rates make it difficult to state this with confidence.
We thus quote the rate of early excesses in SNe Ia as $12.0 \pm 3.6 \%$.

The rate of early excesses in SNe Ia which we measure here is consistent with the rate calculated in \citet{deckers_ztf_excesses} of $18 \pm 11 \%$ 
(calculated from the same dataset),
and is also consistent with the limit of $\lesssim$30\% based on the non-detections in \citet{miller_rise_times},
again calculated from the same dataset.
It is also consistent with the rate calculated in \citet{burke_LCO_sample} of $33\pm19\%$.
\citet{magee_models} 
find 5 excesses in a sample of 23 SNe,
corresponding to a rate of $22\pm10\%$,
which we are also consistent with.
The theoretically expected rate quoted in \citet{kasen} is $\sim$10\%
for strong excesses:
our $5\sigma$ cut should be able to detect such excesses,
and we are again consistent with this value.
The rates of early excesses from all these studies,
with their widely varying methodologies,
are consistent with the null hypothesis that SNe Ia predominantly arise from single-degenerate systems.

Additionally,
as seen in Table \ref{tab:EEx_parameters},
all 11 early-excess SNe are classified as normal SNe Ia.
The sample contains 13 overluminous SNe Ia
(i.e. 99aa-like or 91T-like),
and \citet{jiang_2018} suggested that such overluminous SNe Ia uniformly have early excesses based on six out of six such objects in their sample exhibiting early excesses.
We do not see this effect here:
most of the overluminous objects in this sample do have relatively low early S/N due to their higher redshift,
but two of them
(ZTF18abgmcmv and ZTF18abauprj)
have early $\rm{S/N}>30$
and still have no detectable early excess.
The early-excess objects are not even clustered to the luminous end of normal SNe Ia,
with only 4 of the 11 having $\Delta m_{15}<1.0$
(as measured by \texttt{SNooPy}).

\subsection{Methodology-dependent Early Excess Detections}\label{ssec:EEx_detections}

\begin{deluxetable}{ccccc}
\tablecaption{Which SNe in this sample have early excesses, as identified by different papers}
\tablehead{
\colhead{[\hyperlink{hyper1}{1}]} & 
\colhead{[\hyperlink{hyper2}{2}]} & 
\colhead{[\hyperlink{hyper3}{3}]} & 
\colhead{[\hyperlink{hyper4}{4}]} & 
\colhead{[\hyperlink{hyper5}{5}]}
}
\startdata
        & & aapqwyv &         &         \\ 
        & &         &         & aaqcozd \\ 
        & &         & aaqqoqs &         \\ 
aavrwhu & &         &         & aavrwhu \\ 
        & &         &         & aawjywv \\ 
        & &         &         & aaxsioa \\ 
        & &         & aayjvve &         \\ 
        & &         &         & aazsabq \\ 
        & & abcflnz &         & abcflnz \\ 
        & & abckujq &         &         \\ 
        & & abcrxoj &         &         \\ 
        & &         & abdfazk & abdfazk \\ 
        & &         & abdfwur &         \\ 
        & &         &         & abfhryc \\ 
        & & abgxvra &         &         \\ 
        & &         &         & abimsyv \\ 
        & &         & abpamut &         \\ 
        & &         &         & abssuxz \\ 
abxxssh & & abxxssh & abxxssh & abxxssh \\ 
\enddata
\tablecomments{
The ``ZTF18" has been omitted from each object name to save space.
As is apparent in the table and as discussed in the text,
the statement ``this SN does/does not have an early excess" is methodology-dependent.
References:
\hypertarget{hyper1}{1}: \citet{yao_2019},
\hypertarget{hyper2}{2}: \citet{miller_rise_times},
\hypertarget{hyper3}{3}: \citet{ztf_colors},
\hypertarget{hyper4}{4}: \citet{deckers_ztf_excesses},
\hypertarget{hyper5}{5}: this work.
}
\label{tab:EExs}
\end{deluxetable}

As noted in Section \ref{sec:data_and_sample},
this sample of objects was introduced in a series of three papers,
and an additional paper \citep{deckers_ztf_excesses} examined the same dataset through the lens of Ni models.
Each of these four previous studies, 
and now the one presented here,
have independently looked for objects with early excesses.
Table \ref{tab:EExs} therefore lists the early-excess objects identified by each paper,
each of which approached the identification with a distinct methodology:
\citet{yao_2019} identified two early-excess objects based on photometric comparison to other objects with extremely early data
(more than 17 days before peak);
\citet{miller_rise_times} identified zero early-excess objects,
since all rising lightcurves were consistent with their general power-law fits;
\citet{ztf_colors} identified six objects which had red bumps at early times in their $g-r$ color evolutions;
and \citet{deckers_ztf_excesses} identified six objects with early excesses,
based on quantitative measures of whether they could be fit by any Ni-mixing models in the model grids from \citet{turtls} and \citet{magee_models}.
As the table makes apparent,
each paper (i.e. each methodology) identified mostly mutually exclusive sets of SNe.
As stated above,
we identify 11 objects with early excesses based on their fits to companion shocking models,
adding another mostly-mutually-exclusive set of early-excess SNe Ia from this sample.

Even though they might seem contradictory,
two things are both true:
this work and each of the papers mentioned above have (different) quantitative measures to detect early excesses,
and also they all mostly disagree on which objects have early excesses.
These statements do not invalidate each other -- 
we merely want to stress that even though the final cuts for which objects have early excesses can be quantitative (i.e. objective),
the assumptions undergirding the methodologies introduce (subjective) biases which end up with different results.

\begin{figure}[t!]
\begin{center}
\includegraphics[width=0.47\textwidth]{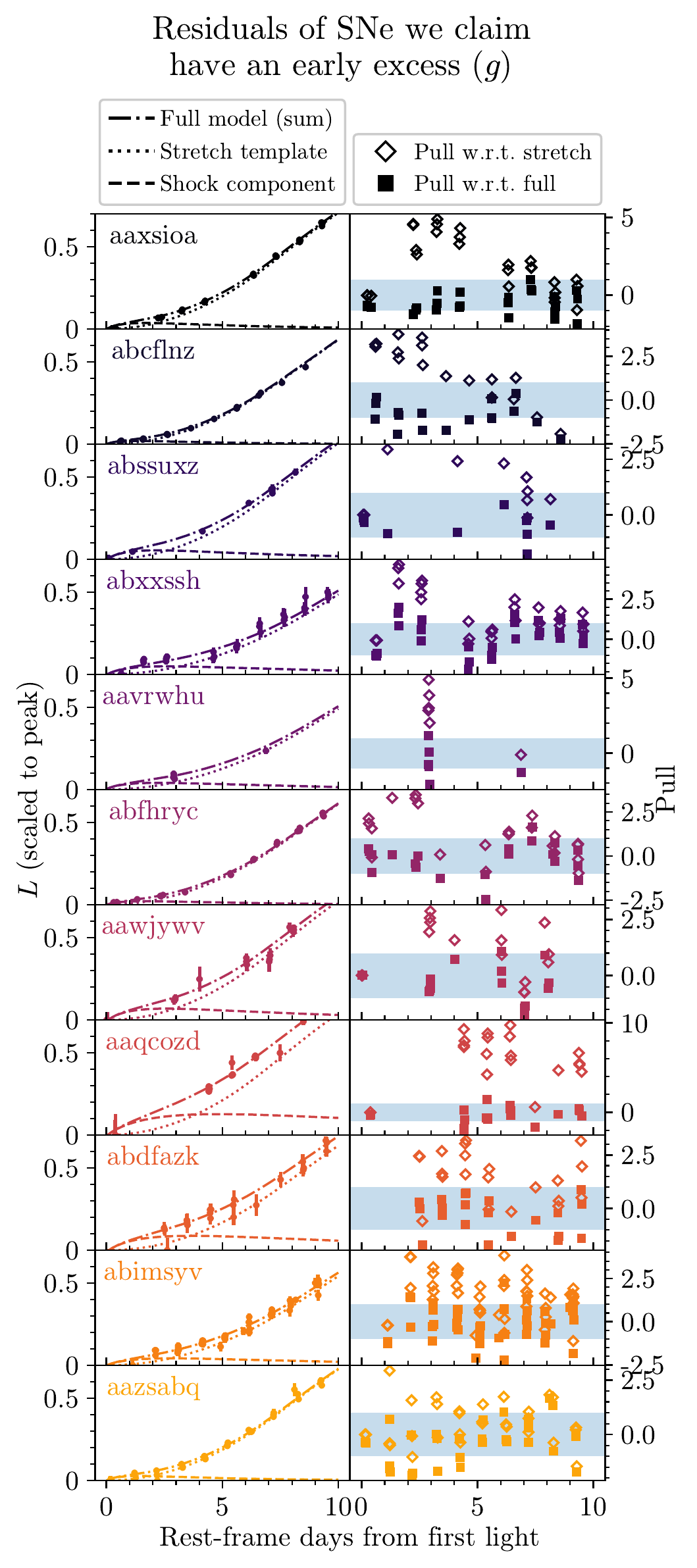}
\caption{
Fits and residuals
for objects we identify as having an early excess.
Objects are ordered by tier (see Table \ref{tab:EEx_parameters}):
the first two SNe are gold-tier
(i.e. the residuals show a strong well-resolved shock component),
the next four are silver-tier
(the shock is not as resolved),
and the last five are bronze-tier.
}
\label{fig:residuals_EEx}
\end{center}
\end{figure}

\begin{figure}[t!]
\begin{center}
\includegraphics[width=0.47\textwidth]{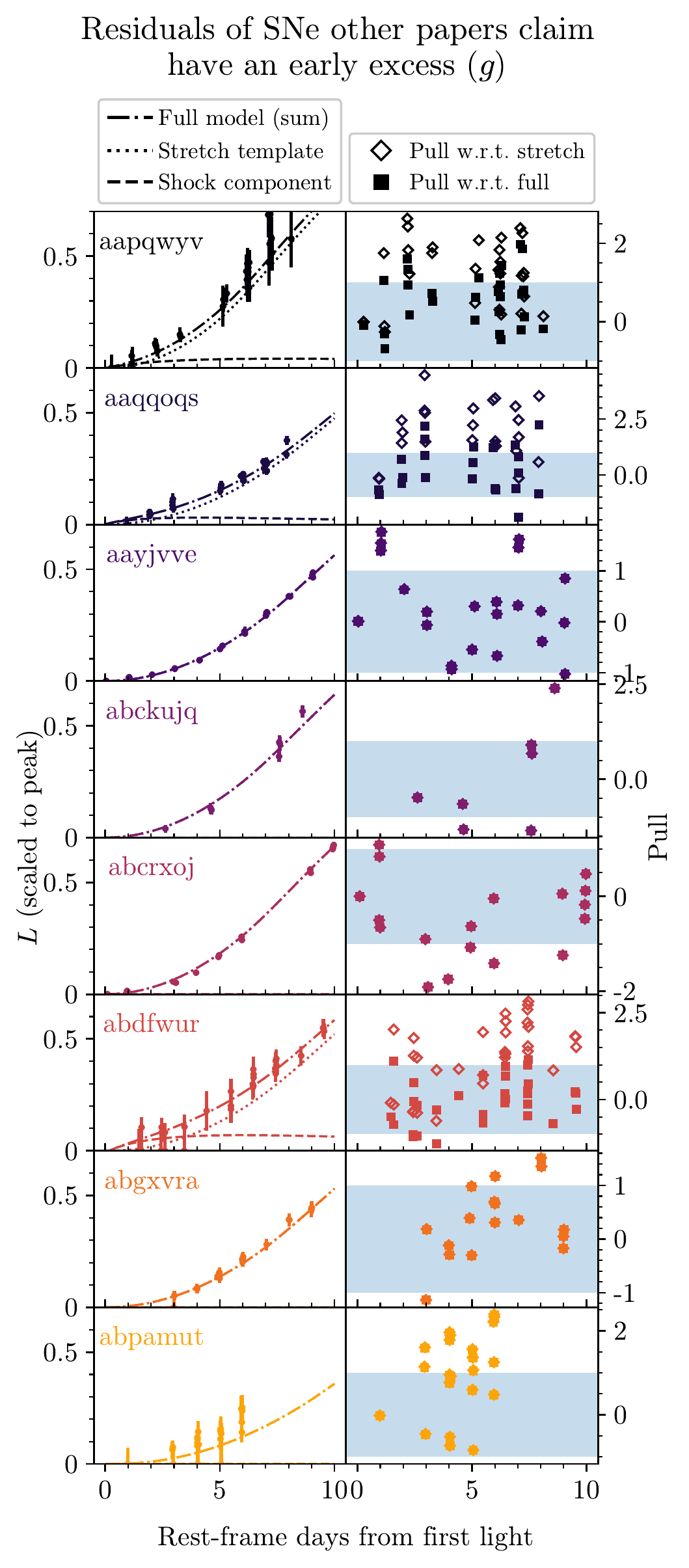}
\caption{
Identical to Figure \ref{fig:residuals_EEx},
but for the objects which other papers identify as having an early excess (see Table \ref{tab:EExs}).
Panels are labelled with the object name, sans ZTF18-.
See the text for discussion of different fits.
}
\label{fig:residuals_not_EEx}
\end{center}
\end{figure}

\begin{figure}[t!]
\begin{center}
\includegraphics[width=0.47\textwidth]{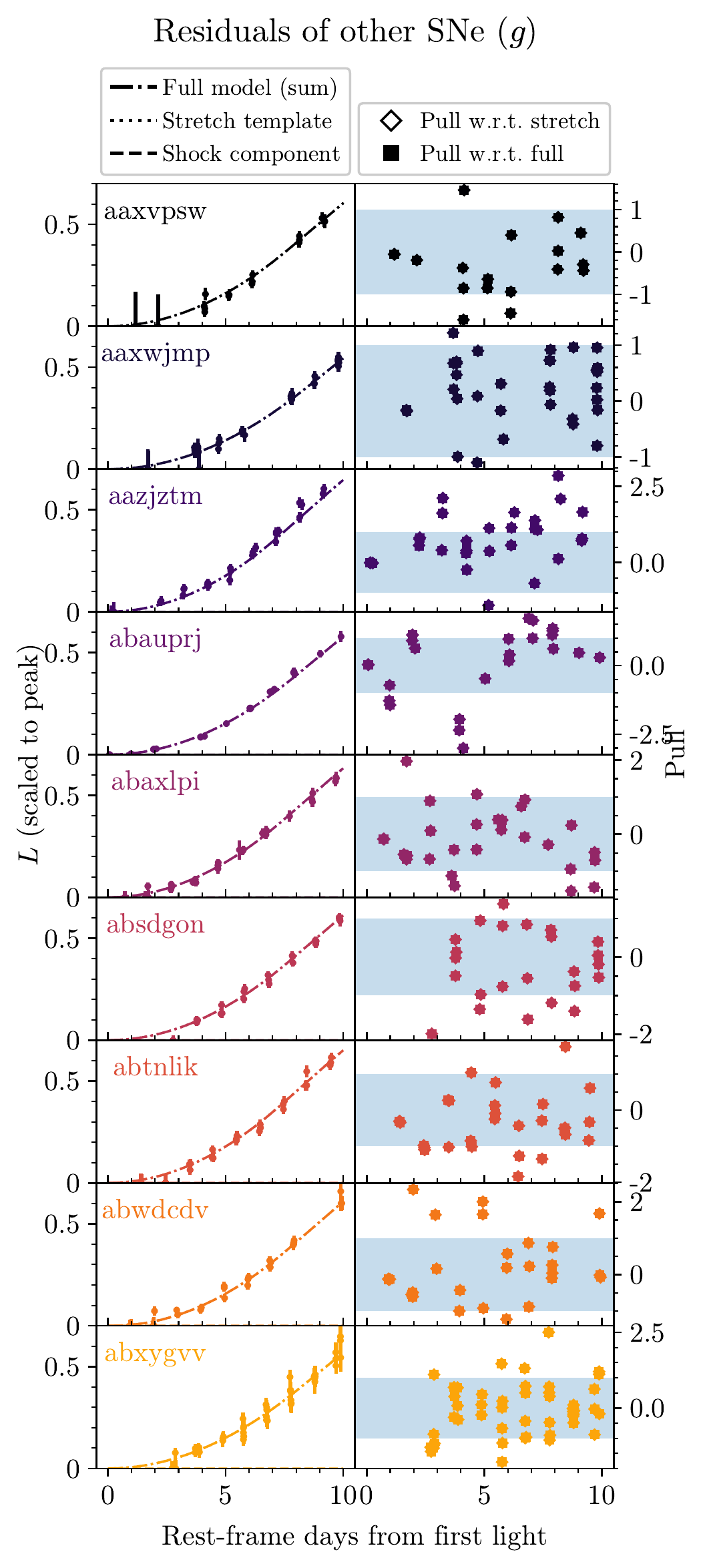}
\caption{
Identical to Figures \ref{fig:residuals_EEx} and \ref{fig:residuals_not_EEx},
but for objects which have especially good early data, and show no signatures of an early excess.
We include this figure to show that the stretch template by itself can be an excellent fit to some objects,
validating its use.
}
\label{fig:residuals_def_not_EEx}
\end{center}
\end{figure}

Figures \ref{fig:residuals_EEx}, \ref{fig:residuals_not_EEx}, and \ref{fig:residuals_def_not_EEx}
show the fits and residuals for three different sets of SNe drawn from the sample.
Figure \ref{fig:residuals_EEx}
shows the 11 objects which we identify as having an early excess,
in the order of their tier as listed in Table \ref{tab:EExs}.
The first two objects (ZTF18aaxsioa and ZTF18abcflnz) are the two ``gold" objects,
because their residuals show a clear and temporally resolved shock component,
manifesting as an initial discrepancy with respect to the stretch template,
which then fades over time.
ZTF18aaxsioa was already plotted in Figure \ref{fig:residuals},
since its residuals have such a clear systematic.
The next four objects 
(ZTF18abssuxz through ZTF18abfhryc)
are in the silver tier,
since either the shock component is not as well resolved temporally (e.g. ZTF18abssuxz)
or only the initial epoch or two are significantly discrepant with the stretch template.
ZTF18abxxssh was classified as an early-excess object by most papers studying this sample (see Table \ref{tab:EExs}),
primarily due to its first two epochs with detections.
The last five objects (ZTF18aawjywv through ZTF18aazsabq) are in the bronze tier:
their residuals don't display systematics as obvious as the above objects,
but nevertheless they are poorly fit by the stretch template alone and the models have unambiguously converged to have some shock component.

The next set of objects (Figure \ref{fig:residuals_not_EEx})
are the eight objects which other papers identify as having an early excess,
but we do not.
Some objects (ZTF18aapqwyv, ZTF18aaqqoqs, ZTF18abdfwur) have converged to a similar region of parameter space as SN 2019np did in \citet{burke_LCO_sample},
with a large best-fit companion separation seen at a high viewing angle needed to explain the discrepancy with the stretch template --
their high viewing angles have excluded them from being listed among our early-excess SNe (see the criteria listed at the start of Section \ref{ssec:EEx_rate}).
This matches the analysis of SN 2019np,
where another study did identify it as having an early excess \citep{2019np_sai},
even though \citet{burke_LCO_sample} did not.
Others (ZTF18aayjvve, ZTF18abgxvra) have no signs of disagreement with our stretch template 
-- in the case of ZTF18abgxvra
only \citet{ztf_colors} identified it as having an early excess,
and since that paper relied on color evolution to identify excesses,
the object's early peculiarities may not be obvious when looking filter-by-filter.

The last set of objects (Figure \ref{fig:residuals_def_not_EEx})
was selected because the objects have exemplary early data,
and also showed no signs of any early excesses
(i.e. the two sets of residuals are on top of each other).
We include this set primarily to show that some objects have early lightcurves which are fit excellently by the stretch template alone,
validating its use.

\begin{figure}[t!]
\begin{center}
\includegraphics[width=0.47\textwidth]{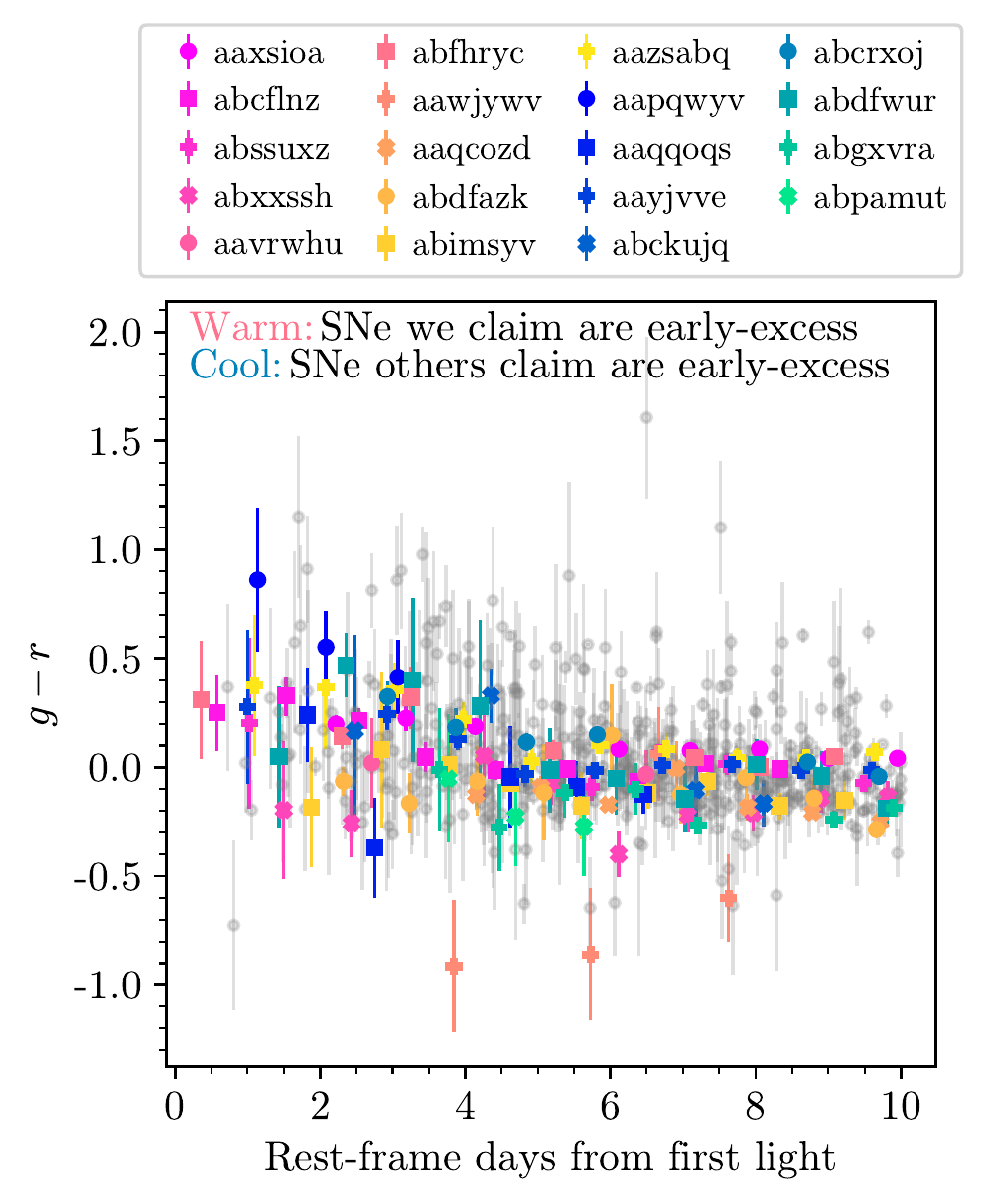}
\caption{
Color evolution of the sample.
The 11 SNe we identify as having an early excess are shown in warm colors,
and the 8 SNe other papers identify as having an early excess are shown in cool colors 
(see Table \ref{tab:EExs} for early-excess classifications).
The rest of the sample is shown in grey.
Colors have been extinction- and K-corrected.
Compare Figures 1 and 4 of \citet{ztf_colors}.
}
\label{fig:color_evolution}
\end{center}
\end{figure}

Lastly,
we show the color evolution of the sample
in Figure \ref{fig:color_evolution},
highlighting early-excess objects.
In \citet{burke_LCO_sample},
the SN with the clearest early excess
(SN 2017cbv)
also had the bluest $g-r$ colors at early times.
That is not quite true here,
where the two SNe with the clearest early excesses in our fits 
(ZTF18aaxsioa and ZTF18abcflnz)
are not the bluest at early times;
however,
ZTF18abxxssh is among the bluest SNe within three days of first light,
and it is the object that most papers identify as having an early excess
(see Table \ref{tab:EExs}).
Generally,
SNe Ia seem to show lower dispersion in their early $g-r$ evolution compared to other colors
\citep[such as $B-V$ or UV filter combinations, see Figures 9 and 10 from][]{burke_LCO_sample},
but we refer to \citet{ztf_colors} for a much more thorough investigation of this sample's color properties.

\section{Conclusion} \label{sec:conclusion}

We have reexamined the 2018 ZTF sample of early SNe Ia \citep{yao_2019, miller_rise_times, ztf_colors},
using models from \citet{kasen} to search for signatures of nondegenerate companion interaction in the form of early excesses in the lightcurves.
We found 11 such objects with signatures of companion interaction:
this na\"{i}vely represents 8.7\% of the sample,
but when compensating for the S/N of the early data
we calculate an overall early excess rate of $12.0\pm3.6\%$.
This rate is consistent with several others calculated throughout the literature,
using different methodologies and/or samples to detect early excesses.
This rate is also consistent with the expectation that SNe Ia predominantly arise from progenitor systems with a Roche-lobe-overflowing nondegenerate companion.
However,
early excesses only occur in this sample in normal SNe Ia and not in any of the 13 overluminous objects,
which is inconsistent with the claim in \citet{jiang_2018} that overluminous SNe Ia have ubiquitous early excesses.

In addition to this result,
we also showed that the detection of early excesses can be methodology-dependent.
For instance,
despite the fact that this same sample had been analyzed by four previous papers,
each of which looked for objects with early excesses,
we identify seven SNe Ia as having early excesses which none of the previous papers identified as such.
This meta-analysis shows that different methodologies make a variety of assumptions about the early lightcurves of SNe Ia,
which can lead to different results.

Companion interaction models have been shown for several years to be excellent (if imperfect) fits to SNe Ia with early excesses
\citep[see e.g.][]{griffin,dimitriadis_18oh,burke_19yvq,griffin_2021aefx,burke_LCO_sample}.
We strongly encourage the observation of large samples of SNe Ia with high-cadence multiwavelength early data
\citep[such as the ones in this paper and in][]{burke_LCO_sample}
so that models' predictions might be tested and the the uncertainty around the progenitor systems of SNe Ia might be gradually reduced.
We also encourage continued theoretical work on
companion interaction models,
since the state of the art is now more than a decade old \citep{kasen}.

\acknowledgments

J.B. and D.A.H. are supported by NSF grants AST-1911151 and AST-1911225, as well as by NASA grant 80NSSC19kf1639.

Time domain research by D.J.S. and G.H. is supported by NSF grants AST-1821987, 1813466, 1908972, \& 2108032, and by the Heising-Simons Foundation under grant \#2020-1864.

\vspace{5mm}
\facilities{Zwicky Transient Facility \citep{bellm_ztf}}

\software{
\texttt{astropy } \citep{2013A&A...558A..33A,astropy},
\texttt{SNooPy} \citep{snoopy},
\texttt{lightcurve\_fitting} \citep{griffin_lightcurvefitting},
\texttt{emcee} \citep{emcee},
\texttt{extinction} \citep{extinction_zenodo}
          }

\bibliographystyle{aasjournal}
\bibliography{biblio}

\end{document}